\documentclass[sigconf,nonacm]{acmart}
\usepackage{algorithm}
\usepackage{float}
\usepackage{enumitem}
\usepackage{subcaption}
\usepackage{algpseudocode}
\providecommand{\Jun}[1]{}

\AtBeginDocument{%
  }

\begin{document}

\title[SARP: An On-SSD Index for KVCache with Fast Prefix Matching and Recovery]{SARP: An On-SSD Index for KVCache with\texorpdfstring{\\}{} Fast Prefix Matching and Recovery}

\author{Jun Kong}
\affiliation{%
  \institution{Arizona State University}
  \country{USA}}
\email{jkong123@asu.edu}

\author{Qi Lin}
\affiliation{%
  \institution{Arizona State University}
  \country{USA}}
\email{qlin36@asu.edu}

\author{Zhichao Cao}
\affiliation{%
  \institution{Arizona State University}
  \country{USA}}
\email{zhichao.cao@asu.edu}

\begin{abstract}
Distributed LLM serving systems rely on a global index to locate reusable key-value cache (KVCache) blocks through prefix matching. If the index is lost after a crash or restart, affected requests cannot locate existing KVCache and recompute it on GPUs. Systems therefore persist the index to restore access to KVCache after a failure. Existing persistent indexes face a tension between prefix-matching performance and recovery time. A snapshot-backed in-memory index supports fast prefix matching, but rebuilding the full index from a snapshot lengthens recovery time. An on-SSD index avoids this rebuild, but adds storage accesses to prefix matching.

Our core insight is that extending a matched prefix over the next short span typically requires distinguishing among only a few candidate token sequences. This suggests that sparse in-memory routing state may suffice to distinguish them. Prefix matching also accesses block metadata in sequence order, allowing metadata for consecutive blocks to be retrieved together. Based on these observations, we present SARP (Sparse Adaptive Routing and Packing), an on-SSD KVCache index for fast recovery and prefix matching. SARP groups consecutive block IDs and metadata into segments to amortize storage accesses across multiple blocks. A compact in-memory routing directory identifies candidate segments using entries at selected block positions. SARP limits segment retrievals and uses compact filters to prioritize candidates with longer potential matches. After retrieval, SARP compares each segment's block IDs with the query's block IDs and reuses only the KVCache blocks covered by the verified prefix. After a restart, SARP restores the sparse routing directory from a snapshot. Across four multi-turn chat and agentic workloads, SARP recovers 8.0$\times$ to 77.5$\times$ faster than the in-memory index baselines and delivers 1.15$\times$ to 66.6$\times$ the request throughput of the on-SSD index baselines, while reusing only 0.0030\% to 0.029\% fewer tokens than exact prefix matching.

\end{abstract}

\maketitle
\renewcommand{\shortauthors}{\shorttitle}

\section{Introduction}
\label{sec:intro}
During large language model (LLM) inference, two requests under the same model and configuration that share a token-level prefix produce the same key and value tensors for that prefix. Serving systems therefore retain these tensors in fixed-length KVCache blocks, so later requests with the same prefix can reuse the corresponding KVCache blocks instead of recomputing them on GPUs~\cite{zheng2024sglang,kwon2023efficient}. This KVCache reuse is common within a chat session because each turn extends the preceding history~\cite{gao2024cost}, and across sessions when requests share system prompts and documents~\cite{li2024scbench}. Claude Code and Codex traces show that 95.7\% of input tokens hit the prefix cache~\cite{zhu2026tracelab}. Because a GPU node's memory cannot hold all reusable KVCache blocks, serving systems offload them to CPU memory, local SSDs, and distributed storage systems~\cite{qin2024mooncake,nvidia-dynamo,alibabaTairKVCache,liu2025lmcache}. A reusable KVCache block can therefore reside on a different node or storage tier from the GPU that needs it.

To locate these blocks across GPU nodes and storage tiers, distributed LLM serving systems maintain a \emph{global KVCache index}~\cite{alibabaTairKVCache,mooncakeStoreDocs,llmdKVIndexerDocs}. The index organizes \emph{block metadata} by token sequence and records where each cached block is available, such as its GPU node and storage tier. It supports two operations. \emph{Prefix matching} compares a request's token sequence with indexed sequences, finds the longest matching prefix whose KVCache blocks are available, and returns that prefix's block metadata in token order. \emph{Prefix insertion} registers a processed sequence with the block metadata of its cached KVCache blocks so later requests can find and reuse them. Existing systems implement these operations with radix-tree or block-hash indexes. SGLang and NVIDIA Dynamo use radix-tree-based indexes, whereas Tair KVCache, Mooncake, vLLM, and llm-d use block-hash indexes~\cite{zheng2024sglang,dynamoRouterDocs,alibabaTairKVCache,qin2024mooncake,vllmPrefixCachingDocs,llmdKVIndexerDocs}. A radix-tree index represents shared prefixes as paths in a compressed tree. A block-hash index maps each block's \emph{chained block ID}, computed from that block and the preceding ID, to its block metadata. The chained ID identifies the prefix that ends at that block.

The in-memory global indexes keep prefix matching fast, but crashes and restarts can make their index state unavailable. When this state is lost, inference requests cannot locate offloaded KVCache blocks and have to recompute their reusable prefixes. Systems preserve index availability in two ways. Replication keeps the index available while a surviving replica remains, as in NVIDIA Dynamo~\cite{dynamoRouterDocs}. Persistence saves the index state so it can be restored later. Mooncake takes snapshots natively~\cite{mooncakeStoreDocs,mooncake1150}, FlexKV uploads its index to Redis~\cite{flexkvDocs}, and Tair KVCache persists its metadata in an external key-value store~\cite{alibabaTairKVCache}.

A snapshot-backed in-memory index is durable, but it cannot serve prefix-matching requests immediately after a restart. Its snapshot stores serialized index state, so recovery has to load the entries and rebuild the DRAM index before serving requests. Recovery time therefore grows with the number of block-metadata entries, and requests cannot locate offloaded KVCache blocks during recovery. \autoref{tab:recovery-cost} reports this recovery time at one billion block-metadata entries: 621 seconds for Redis, approximately 33 minutes for Mooncake, and approximately 52 minutes for Dynamo 1.3. Requests therefore have to recompute their reusable prefixes throughout this recovery time.

An on-SSD index keeps its index representation in SSD-backed durable storage. Once the storage backend has recovered, it can serve prefix-matching requests without rebuilding a full copy in DRAM. This placement shortens recovery time, but prefix matching now retrieves index records from storage and therefore incurs storage-access cost. A key-value store such as RocksDB provides one example of this placement~\cite{rocksdb}: a block-hash index stores one key-value record for each block-hash entry, whereas a radix-tree index stores one record for each tree node. A prefix match may retrieve a chain of records along the shared prefix, so one match can require multiple record retrievals.

\autoref{fig:eval_ssd_placement} shows that the evaluated RocksDB-on-NVMe implementations raise mean prefix-match latency by 39$\times$ to 402$\times$ relative to their DRAM counterparts across the four workloads. A snapshot-backed in-memory index requires rebuilding after a restart, whereas an on-SSD index avoids that rebuild but adds storage-access cost to prefix matching. This tradeoff motivates our research question: \emph{how can a persistent KVCache index reduce both the state rebuilt during recovery and the storage accesses
needed for prefix matching?}

The workloads offer two properties that create opportunities to address these two costs. First, prefix matching follows a query and a cached sequence from their beginnings. Once they agree through a prefix, cached sequences may diverge in different ways. Matching follows the alternative whose tokens agree with the query and accesses block IDs and their corresponding block metadata in sequence order. This sequence order creates an opportunity to retrieve nearby block IDs and metadata together and amortize storage accesses. Second, after a matched prefix reaches the start of a 256-token span, only a few alternatives appear within most spans. This small number of alternatives suggests that less in-memory state may suffice to distinguish them while block IDs and block metadata remain in durable storage. Across the measured spans, counts exceed three in 21.31\% of CodeAct, 0.60\% of LMCache-Agent, 1.78\% of EABF, and 16.85\% of UltraChat.

Using these properties together creates three challenges. First, larger retrieval units amortize storage accesses, but a shared prefix can end inside one unit, so matching only complete units can lose reusable blocks. Retaining routing information for every block preserves such partial matches, but enlarges the in-memory state rebuilt after a restart. Second, spans with many alternatives can require several storage retrievals to determine how far the query matches. A retrieval budget bounds prefix-match cost, but it can leave a longer matching alternative unexamined and reduce KVCache reuse. Third, removing routing information can merge distinct alternatives and increase retrievals. The index therefore has to improve discrimination where many alternatives occur, choose which alternatives to compare under a retrieval budget, and minimize restored routing state without sacrificing prefix reuse.

SARP addresses these challenges with an on-SSD segment store and a sparse in-memory routing directory that maps selected prefixes to candidate segments. The segment store groups consecutive block IDs and their corresponding block metadata in one segment record, so one segment retrieval can cover several blocks. After each retrieval, SARP compares the block IDs with the query before reusing any KVCache blocks. To improve discrimination where candidate lists are large, adaptive routing-point refinement adds routing points at selected block positions and narrows the candidates available to later matches. Candidate filtering uses compact filters to estimate how far each candidate may match, prioritizes candidates with larger estimates, and caps segment retrievals per query window. To reduce restored state, adaptive routing-point pruning removes routing points when the preceding candidate list already fits within the retrieval budget while retaining the candidates needed for exact verification. After a restart, SARP restores only the routing directory, while the segment records remain durable.

Across four multi-turn chat and agentic workloads, SARP recovers 8.0$\times$ to 77.5$\times$ faster than the in-memory index baselines and delivers 1.15$\times$ to 66.6$\times$ the request throughput of the on-SSD index baselines, while reusing only 0.0030\% to 0.029\% fewer tokens than exact prefix matching.

\begin{samepage}
This paper makes three contributions:
\begin{itemize}
    \item We characterize the tension between recovery time and prefix-matching performance in persistent KVCache indexes.
    \item We show that sequence-ordered block metadata and the few alternatives that follow most matched prefixes in multi-turn chat and agentic workloads create opportunities to amortize storage accesses and reduce in-memory matching state.
    \item We present \emph{SARP}, which combines an on-SSD segment store with sparse in-memory routing, routing-point refinement, candidate filtering, and routing-point pruning to reduce recovery state and storage accesses while preserving block-level prefix matching.
\end{itemize}
\end{samepage}

\section{Background}

\subsection{KVCache Reuse and Offloading}\label{sec:kvcache-sharing-background}
During LLM inference, a Transformer-based model processes an input token sequence and
produces key and value tensors for each token~\cite{vaswani2017attention,kwon2023efficient}. Because each token's tensors
depend only on that token and the token sequence before it, the same initial token
sequence produces the same key and value tensors across inference requests with the same
model and configuration~\cite{kwon2023efficient}. When two inference requests share a number of tokens from the beginning, those common tokens form a shared prefix, and reusing the
corresponding key and value tensors would avoid GPU recomputation. Serving systems therefore retain
key and value tensors in a \emph{KVCache}. Mainstream inference systems (e.g., vLLM) organize them into \emph{KVCache
blocks}, each holding the tensors for a fixed number of consecutive tokens so
later inference requests can reuse these blocks
~\cite{kwon2023efficient,zheng2024sglang}.

These shared prefixes can recur within a \emph{session}, or a sequence of related
inference requests during an ongoing conversation or agent interaction.
Successive requests within a session often extend the same history, so their
shared prefixes with earlier requests grow over time. In a conversational
session, each new request includes earlier
questions and responses followed by the user's latest message. In ShareGPT
conversations, recomputing KVCache for earlier turns can account for up to 99\%
of the time spent processing the input~\cite{gao2024cost}. Within an agent
session, tool results are similarly appended to the history before the next
inference request~\cite{zhu2026tracelab}. For example, TraceLab reports that 95.7\% of input tokens in
Claude Code and Codex traces hit the prefix cache~\cite{zhu2026tracelab}.
Across sessions, independent inference requests can also begin with a shared
prefix, such as system prompts~\cite{li2024scbench,claudePromptCaching,microsoftPromptCaching}.
By avoiding repeated computation over these shared prefixes,
KVCache reuse can effectively reduce serving cost and Time To First Token (TTFT)
~\cite{wang2025kvcache}.

Realizing these reuse opportunities requires retaining KVCache blocks, but a GPU
node's HBM alone cannot hold all of them as workloads grow
~\cite{qin2024mooncake}. Serving systems
therefore offload KVCache to CPU memory, local SSDs, and even shared storage
backends such as object stores and distributed file systems. Mooncake
distributes paged KVCache blocks across nodes and transfers them through RDMA
~\cite{qin2024mooncake}. NVIDIA Dynamo manages KVCache across GPU, CPU,
local-disk, and object-storage tiers~\cite{nvidia-dynamo}. Tair KVCache~\cite{alibabaTairKVCache} and
LMCache~\cite{liu2025lmcache} provide interfaces for storing KVCache in
heterogeneous backends. Across these systems, retained
KVCache blocks can reside on different nodes and storage tiers.

\subsection{Global Index for KVCache} \label{sec:kvcache-sharing-index}
To locate and manage reusable KVCache blocks across GPU nodes and storage
tiers, serving systems use a global index as the core metadata structure
~\cite{alibabaTairKVCache,mooncakeStoreDocs,llmdKVIndexerDocs}.
Such an index organizes \emph{block metadata} by token sequence,
associating each sequence with metadata for its cached KVCache
blocks.
This block metadata records where each block is available, such as its GPU node
and storage tier.

To find reusable blocks and register newly cached ones, the global index usually supports two index
operations. \emph{Prefix matching} compares a new request's token sequence
with sequences indexed by the global index and finds the longest matching prefix
whose KVCache blocks are available. It returns the block metadata of the \emph{matched prefix} in token order. \emph{Prefix insertion} registers a
processed token sequence together with the block metadata of its cached KVCache
blocks, so later requests can find and reuse those KVCache blocks. Existing
systems use indexes based on radix trees or block-hash tables. SGLang and
NVIDIA Dynamo use
radix-tree-based indexes, whereas Tair KVCache, Mooncake, vLLM, and llm-d use
block-hash indexes
~\cite{zheng2024sglang,dynamoRouterDocs,alibabaTairKVCache,qin2024mooncake,vllmPrefixCachingDocs,llmdKVIndexerDocs}.

\textbf{Radix-tree index.}
A radix-tree index represents shared prefixes as paths from the tree root, so
sequences with the same prefix share the corresponding path
~\cite{morrison1968patricia,leis2013art}. Each edge
represents a consecutive group of tokens or block IDs. During prefix matching,
the index compares the request with these edges until the request ends or the
next edge differs. The traversed path then identifies the matched prefix and
its associated block metadata~\cite{zheng2024sglang,dynamoRouterDocs}.

\textbf{Block-hash index.}
A block-hash index divides each sequence it indexes into fixed-size blocks
$B_1,B_2,\ldots$ and computes a chained block ID
$h_i = hash(h_{i-1}, B_i)$,
where $h_0$ is a fixed initial value. Because each ID includes the preceding
ID, $h_i$ identifies the block-aligned prefix ending at $B_i$, apart from hash
collisions. The index maps each $h_i$ to the block metadata for $B_i$. During
prefix matching, it computes the request's chained block IDs and probes them from the first
block until the first miss~\cite{vllmPrefixCachingDocs}. Successful probes identify the blocks in the
matched prefix and return their block metadata
~\cite{alibabaTairKVCache,qin2024mooncake,vllmPrefixCachingDocs,llmdKVIndexerDocs}.

\subsection{Index Persistence and Recovery}
Existing systems generally keep the global index in memory for fast prefix
matching, but this placement leaves the index state volatile across crashes
and restarts. A crash or restart can make otherwise intact KVCache blocks
undiscoverable, forcing inference requests to recompute them
~\cite{mooncake1150} and increasing TTFT. To avoid this loss of index availability, index replication
can keep the index available while a surviving replica remains, as in NVIDIA
Dynamo~\cite{dynamoRouterDocs}. Beyond replication, some index systems also support \emph{persistent indexes} that preserve the index state after restart by snapshot or offloading indexes to durable storage.

A \emph{snapshot-backed in-memory index} periodically saves a durable snapshot of the whole in-memory index state.
Mooncake Master Service performs this snapshotting directly
~\cite{mooncakeStoreDocs}. llm-d can instead keep its index in Redis
~\cite{llmdKVIndexerDocs}, an in-memory store that persists its data through
snapshots~\cite{redisPersistenceDocs}. After a restart, the system loads the snapshot and
rebuilds the in-memory index before serving prefix-matching requests
~\cite{redisPersistenceDocs}.

An alternative, \emph{on-SSD index}, keeps the index representation itself in
SSD-backed durable storage. A persistent key-value store or database such as RocksDB
~\cite{rocksdb} is one possible backend for this placement and can store either
index structure as key-value records: a block-hash index maps each block ID to
its block metadata, whereas a radix-tree index stores each node keyed by its
node ID. After a restart, the on-SSD index opens its storage backend and
immediately serves queries.

\section{Motivation}\label{sec:motivation}

\subsection{The Persistence and Performance Tension} \label{sec:dillema}

Persistent KVCache indexes face a tension between prefix-matching performance
and recovery time. In-memory indexes provide fast prefix matching but recover
slowly after a restart. On-SSD indexes recover quickly but add slow storage accesses
to the prefix-matching path.

For an in-memory index with snapshot, recovery rebuilds the index from a snapshot after restart. During recovery, the index cannot serve prefix-matching queries, leaving offloaded KVCache undiscoverable. \autoref{tab:recovery-cost} reports the recovery time of each index, from loading its snapshot until it can serve prefix-matching queries, at one billion block-metadata entries. At this scale, our experiments report 621 seconds for Redis and estimate approximately 33 and 52 minutes for Mooncake and Dynamo 1.3, respectively. These recovery times can leave existing KVCache undiscoverable for minutes after a restart.

\begin{table}[tbp]
    \centering
    \small
    \setlength{\tabcolsep}{4pt}
    \caption{Index recovery at one billion block-metadata entries.}
    \label{tab:recovery-cost}
    \begin{tabular}{@{}llr@{}}
        \toprule
        System & Index placement & Recovery time \\
        \midrule
        Redis 8.10   & In-memory & 621~s \\
        Mooncake 0.3 & In-memory & $\sim33$~min \\
        Dynamo 1.3   & In-memory & $\sim52$~min \\
        RocksDB      & On-SSD    & $\sim100$~ms \\
        \bottomrule
    \end{tabular}
\end{table}

\begin{figure}[tbp]
    \centering
    \includegraphics[width=\columnwidth]{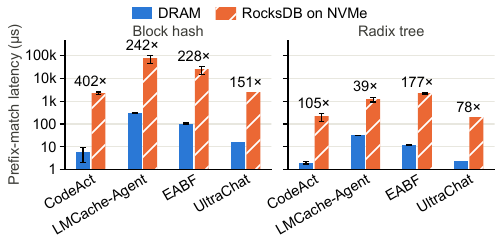}
    \caption{Mean prefix-match latency for the block-hash index and radix tree
    in DRAM and in RocksDB on NVMe. RocksDB uses a per-workload block cache sized at 10\% of its storage files. Labels give the RocksDB-to-DRAM latency
    ratio.}
    \label{fig:eval_ssd_placement}
    \Description{Two bar charts share a log-scale y-axis of mean prefix-match
    latency in microseconds: block hash on the left and radix tree on the
    right. Each has four dataset groups, CodeAct, LMCache-Agent, EABF, and
    UltraChat, with a solid blue bar for DRAM and a hatched orange bar
    for RocksDB on NVMe with a 10 percent block cache. RocksDB is slower in
    every group. The slowdown is 402, 242, 228, and 151 times for the block
    hash, and 105, 39, 177, and 78 times for the radix tree.}
\end{figure}

To reduce the recovery time, an on-SSD index stores the index itself on durable storage. After a
restart, the storage engine reopens its durable files and rebuilds only its
own metadata. The index can then serve prefix-matching queries without
rebuilding a separate full copy in DRAM.\@ RocksDB~\cite{rocksdb} provides one
concrete example of this on-SSD placement. \autoref{tab:recovery-cost} shows
that RocksDB recovers in approximately 100~ms at this scale. This faster recovery comes with a
query-path cost: prefix matching retrieves records from SSD, and those accesses
can slow down the query. \autoref{fig:eval_ssd_placement} shows that, across
the four workloads in \autoref{tab:workloads}, mean prefix-match latency in RocksDB on NVMe is
151$\times$ to 402$\times$ the corresponding DRAM latency for the in-memory
block-hash index and 39$\times$ to 177$\times$ for the radix tree.

\begin{table}[tbp]
    \centering
    \small
    \setlength{\tabcolsep}{4pt}
    \caption{Dataset characteristics.}
    \label{tab:workloads}
    \begin{tabular}{@{}llrrr@{}}
        \toprule
        Dataset & Type & Sessions & Turns
            & \begin{tabular}[b]{@{}r@{}}Tokens per session\end{tabular} \\
        \midrule
        CodeAct       & Agent & 7.1K & 59.4K & 1.5K  \\
        LMCache-Agent & Agent & 767  & 24.9K & 42.0K \\
        EABF          & Chat  & 3.5K & 198K  & 14.2K \\
        UltraChat     & Chat  & 487K & 2.82M & 1.2K  \\
        \bottomrule
    \end{tabular}
\end{table}

These observations set our design goal: recover the persistent index quickly
after a restart while preserving high prefix-matching performance.
This goal motivates our central question: \textbf{\emph{how can a persistent
KVCache index reduce both the state rebuilt during recovery and the storage
accesses needed for prefix matching?}}

\subsection{Insights and Observations}\label{sec:opportunities}

Persistent KVCache indexes need to recover quickly while keeping prefix matching fast. Our key insight is that the workload properties of LLM prefix matching provide opportunities to amortize storage accesses and reduce in-memory state.

During prefix matching, given a query and cached token sequences, prefix matching compares them from their beginnings and progressively extends the matched prefix. Multiple cached token sequences can share the current matched prefix and differ later. The index distinguishes these alternatives, and prefix matching follows the alternatives whose tokens continue to match the query.

\textbf{Coarse retrieval opportunity.} As prefix matching advances along a cached token sequence, it accesses block metadata in sequence order. Co-locating this metadata could let one retrieval load it together, amortizing storage accesses.

\textbf{Sparse in-memory state opportunity.}
Consider a fixed-length span immediately following a matched prefix.
For most spans, only a few alternatives follow the matched prefix within the span.
Thus, as matching advances over the span, the index needs to distinguish only those few alternatives.
This small number of alternatives suggests that less in-memory state may suffice to guide prefix matching.

To quantify this observation, we partition each session history into consecutive, non-overlapping spans of 256 tokens. The final span may be shorter. For each span, we count how many alternatives can follow a prefix match that reaches the span's start. Across all measured spans, counts exceed three in 21.31\% of CodeAct spans, 0.60\% of LMCache-Agent spans, 1.78\% of EABF spans, and 16.85\% of UltraChat spans. Thus, most spans contain at most three alternatives, while a minority contain more. These counts indicate that a persistent index can keep only a small amount of routing state for most spans and retain the remaining block metadata on storage.

Amortizing storage accesses and reducing in-memory state expose three
challenges.

\textbf{Challenge 1: Storage organization.}
How should storage place block metadata to support prefix insertion and
efficient retrieval and comparison?

\textbf{Challenge 2: Retrieval count.}
After a matched prefix, several cached token sequences can share that prefix
and then diverge into different alternatives.
The query follows the alternative whose tokens match, so the
index may need to retrieve and compare multiple alternatives to determine how
far the match extends.
How can the index bound these retrievals without losing too much KVCache reuse?

\textbf{Challenge 3: In-memory state size.}
Removing in-memory state can leave distinct alternatives indistinguishable.
How can the in-memory state stay small while preserving the distinctions
needed for prefix matching?

\section{Design} \label{sec:design}

\subsection{Overall Architecture}
To address the challenges presented in \autoref{sec:opportunities}, we propose \textbf{\emph{SARP}} (Sparse Adaptive Routing and Packing), an on-SSD index for large-scale KVCache. It combines an on-SSD segment store, which bounds and amortizes storage accesses during prefix matching, with a sparse in-memory routing directory, which keeps the state restored after a restart small.

\begin{figure}[tbp]
    \centering
    \includegraphics[width=\columnwidth]{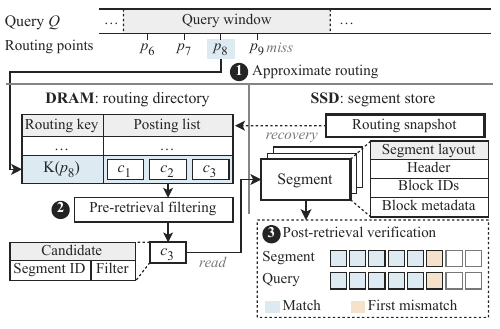}
    \caption{SARP architecture.}    \label{fig:architecture}
    \Description{A query token sequence at the top shows its current query
    window, marked with routing points p6 to p9. The routing key at p9 misses,
    and approximate routing uses the key at p8 to select its entry in the
    routing directory in DRAM, which maps routing keys to posting lists, here
    candidates c1 to c3. Each candidate holds a
    segment ID and a filter. Pre-retrieval filtering keeps c3, whose segment
    is read from the segment store on SSD. A segment holds a header, block
    IDs, and block metadata. Post-retrieval verification compares the segment
    with the query block by block: five blocks match and the sixth is the
    first mismatch. A routing snapshot on SSD restores the routing
    directory during recovery.}
\end{figure}

\textbf{Architecture.}
The overall architecture of SARP is shown in \autoref{fig:architecture}.
The in-memory \emph{routing directory} locates candidate segments during
prefix matching, while the on-SSD
\emph{segment store} retains the segments used for exact verification. Each segment stores the block IDs and block metadata of up to \(S\) consecutive blocks of one cached sequence in sequence order. A later sequence that matches a whole segment, diverging only after the segment ends, reuses that segment, so SARP copies only the segment in which two sequences diverge. Once a segment is located, one segment retrieval returns the block metadata of up to \(S\) consecutive blocks, whereas existing on-SSD indexes retrieve one record per block or tree node.

To locate these segments without SSD I/O, the in-memory \emph{routing directory} keeps
\emph{routing entries} at selected block positions of each segment, called \emph{routing points}. In this way, we do not need to index each KVCache block in memory, which effectively reduces memory utilization.
Each entry of the routing directory maps a \emph{routing key} identifying the prefix through its
routing point to a \emph{posting list} of references to segments sharing
that prefix. The posting list obtained during a prefix match forms its
\emph{candidate set}.
A segment can have several routing points, whose spacing is independent
of segment size.

\textbf{Matching path.}
SARP matches the query from the beginning, processing one query window of
up to \(S\) blocks at each step.
(1)~\emph{Approximate routing.} SARP probes the routing key for the
query prefix at the deepest routing point reached in the current query window.
If that key misses, it falls back to the previous available routing point and
obtains that point's posting list of segment references.
(2)~\emph{Pre-retrieval filtering.} Before retrieving any segment, SARP computes the query's filter
and compares it with each candidate reference's filter to rank candidates by a
conservative filter-derived match potential. It performs at most
\texttt{MaxRetrieval} segment retrievals per window to bound storage access
(\autoref{sec:candidate-filtering}).
(3)~\emph{Post-retrieval verification.} SARP compares the block IDs in
each retrieved segment with the current query window's block IDs before
reusing any blocks.

\textbf{Insertion path.}
Prefix insertion uses the matching path to identify a matched segment, then
compares its block IDs with the new sequence's block IDs to find their shared
prefix.
If the shared prefix reaches the end of the stored segment,
SARP extends the segment in place with the unmatched suffix, and an empty
suffix requires no write. Otherwise, if the shared prefix ends before the segment end, SARP creates a new segment, copies the shared prefix, and appends the unmatched suffix. After either operation, it registers the segment in the in-memory routing directory.

\textbf{Recovery.}
To locate these persistent segments after a restart, SARP restores
the routing directory, including candidate references and filters, from
its snapshot. Once the segment store has recovered, segment
records become queryable without rebuilding a full per-block index in
application DRAM.

\subsection{Adaptive Routing-Point Refinement}
\label{sec:routing-point-placement}
A routing entry at a selected routing point stores a posting list of references
to segments that share the prefix through that point.
When many such segments differ later in the same interval, the posting list
becomes large.
A large posting list can require prefix matching to consider more segment
references and can increase segment retrievals. Under a retrieval cap, it may
also leave a longer matching segment unexamined.
Adding another routing point can provide a smaller posting list to prefix
matches that reach it, but it enlarges the in-memory routing state restored
during recovery.

SARP therefore starts with a small set of evenly spaced routing points
and refines only intervals whose posting-list size exceeds a fixed threshold
\(\tau\). This policy adds discrimination where it is needed while avoiding
unnecessary routing state and workload-specific profiling or tuning.

\textbf{Routing entries.}
A routing point at block position \(p\) registers its segment under
the key \(K(p)\).
Segments with the same key share a routing entry
\(K(p) \mapsto \mathcal{C}(p)\), where \(\mathcal{C}(p)\) is the posting
list of references to segments carrying that key.
The routing key is \emph{prefix-causal}: it depends only on \(p\) and the
prefix through \(p\), so later blocks cannot change the entry for that prefix.

SARP satisfies this requirement by using the block ID at \(p\) as
\(K(p)\). Chained hashing makes this ID depend on every block through \(p\),
but on no block after \(p\).

\textbf{Base placement.}
Consider a segment of \(S\) blocks (\(S=64\) by default) beginning at
position \(s\), with exclusive end \(e=s+S\).
SARP places \(R\) fixed routing points (\(R=4\) by default) at

$$
    p_i = s + \left\lfloor \frac{iS}{R} \right\rfloor,
    \qquad 0 \le i < R .
$$

This placement partitions the segment into \(R\) consecutive intervals.
The first point \(p_0=s\) gives every prefix match an initial routing entry.
The exclusive end \(e\) needs no routing entry because no block follows it.

Let \(I_i=[p_i,p_{i+1})\), with \(p_R=e\), denote a fixed routing interval.
For a query whose prefix match ends at \(q\in I_i\), \(p_i\) is the latest
fixed routing point available to that prefix.

The key \(K(p_i)\) captures only the prefix through \(p_i\), so it cannot
distinguish segments that share this prefix but differ later in \(I_i\).
These segment references therefore remain grouped in the posting list
\(\mathcal{C}(p_i)\). SARP refines \(I_i\) when this posting list
exceeds the fixed threshold \(\tau\) by placing one additional routing point:
$$
    |\mathcal{C}(p_i)| > \tau ,
$$

\textbf{Refinement tradeoff.}
The position of an additional routing point trades reachability against
discrimination.
An earlier point is available to more prefix matches, but its key covers a
shorter prefix and leaves more segment references grouped together.
A later point uses more prefix information and can distinguish more segments,
but only matches that reach that position can use it.

\begin{figure}[tbp]
    \centering
    \includegraphics[width=\columnwidth]
        {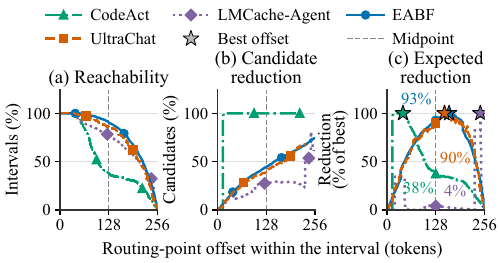}
    \caption{Routing-point placement: reachability, candidate reduction, and
    expected candidate reduction relative to the best offset.}
    \label{fig:fig_design1_placement}
    \Description{Three line charts side by side, each with CodeAct,
    LMCache-Agent, EABF, and UltraChat against a routing point position from 0 to
    256 tokens. Reachability falls from 100 to 0 percent as the position
    moves later. At the midpoint it is about 38 percent for CodeAct, 77 percent
    for LMCache-Agent, 90 percent for EABF, and 85 percent for UltraChat.
    Candidate reduction rises steadily for EABF and UltraChat, to about 45 and 41
    percent at the midpoint, while for CodeAct it jumps to nearly 100
    percent at the 14th token, and for LMCache-Agent it stays below 30
    percent until the 232nd token and then rises to about 53 percent. The
    expected candidate reduction peaks at offsets of 42 tokens for CodeAct,
    245 tokens for LMCache-Agent, and 163 and 151 tokens for EABF and
    UltraChat.}
\end{figure}

\autoref{fig:fig_design1_placement} measures this tradeoff.
Each observation pairs a request with one fixed 256-token interval
\(I_i=[p_i,p_{i+1})\) that starts at token 256 or later.
We skip the first interval because its initial posting list contains every
segment.
For each remaining interval, \(\mathcal{C}(p_i)\) is the posting list of
segment references that share the request's prefix through \(p_i\).
We retain observations whose longest prefix match ends inside the interval.
Among these observations, we keep the 10\% with the largest posting lists to
focus on the intervals targeted by refinement.
For an additional point at offset \(k\), \emph{reachability} is the share of
observations whose prefix match reaches \(k\).
We measure discrimination as \emph{candidate reduction}, the share of segment
references in \(\mathcal{C}(p_i)\) that the point rules out for those
observations.
\emph{Expected candidate reduction} combines both as the mean number of
references the point rules out per observation, counting zero for
observations that do not reach \(k\).
As the offset of the additional point grows, reachability falls
(\autoref{fig:fig_design1_placement}a) and candidate reduction rises
(\autoref{fig:fig_design1_placement}b).
Expected candidate reduction (\autoref{fig:fig_design1_placement}c) peaks near
the middle of the interval on EABF and UltraChat, at 163 and 151 tokens,
while on the agent workloads it peaks near an end of the interval.
On CodeAct, it peaks at only 42 tokens, because 93\% of the segment
references stop matching at the 14th token.
On LMCache-Agent, it peaks at 245 tokens, because candidate reduction stays
below 30\% until the 232nd token and then rises to 53\%.
A fixed position selected for one workload may therefore be less effective on
another.

\textbf{Midpoint and last-block refinement.}
Because the best offset varies across workloads, SARP uses the
midpoint as the best fixed default when no workload profile is available.
It balances reachability and discrimination without workload-specific
placement.
At the midpoint, the expected candidate reduction in
\autoref{fig:fig_design1_placement}c reaches 38\%, 4\%, 93\%, and 90\%
of the best offset on CodeAct, LMCache-Agent, EABF, and UltraChat.
When prefix matches stop uniformly across an interval, the midpoint is the
best single position because it splits the interval evenly.
When match endings are skewed toward one side, a prefix match either reaches
the midpoint and can use the added entry, or ends before it and continues with
the fixed entry.
The fixed entry therefore remains available for earlier matches, without
requiring workload-specific placement.
For an interval selected for refinement, \(I=[p_\ell,p_r)\), the additional
routing point is placed at
\[
    m(I) =
    p_\ell +
    \left\lfloor \frac{p_r-p_\ell}{2} \right\rfloor .
\]

Segments in \(\mathcal{C}(p_\ell)\) that cover \(m(I)\) are grouped by
their block ID at that position, and each group is registered under the
corresponding routing key.
Only the segment's final interval receives an additional last-block routing
point. If the midpoint posting list in that interval still exceeds \(\tau\),
SARP places this point at \(e-1\), the segment's last block.
This last-block probe handles segments that still differ near the segment end.
With \(S=64\) and \(R=4\),
the final interval is \([s+48,s+64)\), with midpoint \(s+56\) and last block
\(s+63\).

\textbf{Materialization and prefix matching.}
When insertion raises \(|\mathcal{C}(p_\ell)|\) above \(\tau\), SARP
materializes midpoint entries for existing segments in \(\mathcal{C}(p_\ell)\)
that cover \(m(I)\).
Subsequent inserted segments that cover the midpoint register their midpoint
keys as well. Segments in the final interval also register the last-block key
when that entry exists.
For the final interval, if the midpoint posting list still exceeds \(\tau\),
SARP also materializes the entry at the segment's last block.
Back-fill of entries for existing segments runs asynchronously in a background
worker, so insertion does not wait for complete materialization.
Until the added entries are complete, the fixed entry at \(p_\ell\) remains
authoritative.
After materialization, a query whose prefix reaches \(m(I)\) probes the
midpoint key.
On a hit, prefix matching uses the midpoint posting list. On a miss, it falls
back to \(\mathcal{C}(p_\ell)\).
A query whose prefix ends before \(m(I)\) uses \(\mathcal{C}(p_\ell)\)
directly.
In the final interval, a query that reaches \(e-1\) probes the last-block
entry when it exists.
On a hit, it uses the last-block posting list.
On a miss, or when the entry does not exist, it uses the deepest earlier
routing entry reached by the query.

\textbf{Placement overhead.}
SARP does not recursively refine the intervals created by a
midpoint. Each of the \(R\) intervals receives at most one midpoint routing
point, and only the final interval can receive one additional last-block
routing point. Thus, a segment contributes at most \(2R+1\) routing-point
registrations. With \(R=4\), this bound is nine registrations per segment.
This bound limits refinement's added routing state and recovery work to a fixed
amount per segment.

\subsection{Candidate Filtering}
\label{sec:candidate-filtering}

Adaptive routing-point refinement reduces the posting list returned by a routing key, but the list may still contain multiple candidate segments.
Verifying every remaining candidate is expensive because each verification incurs a random SSD read.
SARP therefore associates each candidate with a \emph{compact filter} that bounds how far the candidate may continue matching the query.
We call this filter-derived upper bound the candidate's \emph{match potential}.

Candidate filtering performs the pre-retrieval filtering step in
\autoref{fig:architecture}. It uses the match potential to choose which candidate segments
to load for exact block-ID verification, reducing segment retrievals when a posting
list is large. The filter layout and search procedure below show how SARP
makes this selection from compact fingerprints.

\begin{figure}[tbp]
    \centering
    \includegraphics[width=\columnwidth]{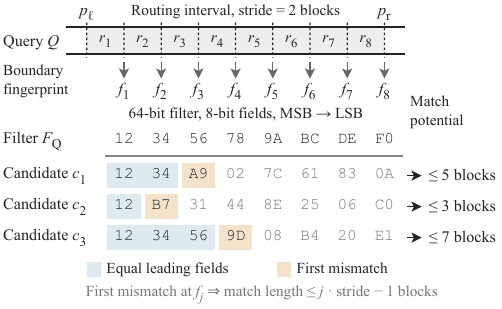}
    \caption{Filter layout and match potential for eight regions of two blocks.}
    \label{fig:filter-layout}
    \Description{The query sequence between routing point p-ell and the
    next routing point p-r is split into eight regions, r1 to r8, of two
    blocks each. The fingerprint of the block IDs in each region gives filter
    fields f1 to f8. The query filter holds the 8-bit fields
    12, 34, 56, 78, 9A, BC, DE, and F0. Three candidate filters are compared
    with it field by field. Candidate c1 first differs at field 3 (A9), so
    its match potential is at most 5 blocks. Candidate c2 first differs at
    field 2 (B7), at most 3 blocks. Candidate c3 first differs at field 4
    (9D), at most 7 blocks. Fields after the first mismatch are shown in
    gray because they are not used.}
\end{figure}

\textbf{Filter layout.}
To encode match potential within a fixed 64-bit budget, SARP divides the routing interval that starts at each routing point into \(N\) consecutive regions of \(\mathit{stride}\) blocks.
For a routing point at position \(p\), region \(r_i\) covers blocks \(p+(i-1)\cdot\mathit{stride}\) to \(p+i\cdot\mathit{stride}-1\), so the first region starts at \(p\).
Let \(\textsc{BlockIDs}(r_i)\) denote the block IDs in region \(r_i\).
The \(i\)-th filter field keeps the low \(b\) bits of a hash \(H\) over them:

$$
f_i =
\textsc{LowBits}_b
\left(
H\left(\textsc{BlockIDs}(r_i)\right)
\right).
$$

Chained hashing makes the last block ID of a region depend on every block through the region end.
Each field is therefore a fingerprint of the prefix through the end of its region.

The filter provides a conservative upper bound on a candidate's true match
length. If a candidate matches the query through a region, the candidate and
query have the same block IDs and therefore the same fingerprint.

The \(N\) region fingerprints form a single 64-bit filter:

$$
F=f_1\|f_2\|\cdots\|f_N,
\qquad b=64/N.
$$

Fields are packed in sequence order from the most significant to the least significant bits, as illustrated in \autoref{fig:filter-layout}.
Unused fields of a partially filled segment are zero-padded.
A matching zero-padded field or a fingerprint collision can delay the first
mismatch and overestimate the candidate's match potential. Exact block-ID
verification after retrieval prevents such an overestimate from admitting
incorrect KVCache reuse.

For a routing interval with \(N\) regions of \(\mathit{stride}\) blocks, the
maximum match length in the routing interval is
\(L_{\max}=N\cdot\mathit{stride}\).

\begin{algorithm}[t]
\caption{\textsc{FindMaxPotential}}
\label{alg:find-max-potential}
\begin{algorithmic}[1]
\Require Query filter $F_Q$, nonempty posting list $P$ sorted by filter,
         maximum match length in the routing interval $L_{\max}$
\Ensure Maximum filter-derived match potential

\Function{FindMaxPotential}{$P,F_Q,L_{\max}$}
    \State $\ell \gets 0,\; r \gets L_{\max}$
    \While{$\ell < r$}
        \State $L \gets \lfloor(\ell+r+1)/2\rfloor$
        \If{\Call{ExistsMatch}{$P,F_Q,L$}}
            \State $\ell \gets L$
        \Else
            \State $r \gets L-1$
        \EndIf
    \EndWhile
    \State \Return $\ell$
\EndFunction

\Function{ExistsMatch}{$P,F_Q,L$}
    \State $k \gets \lfloor L/\mathit{stride}\rfloor$
    \State $M_L \gets \Call{HighMask}{k\cdot b}$
    \State $F_{\min} \gets F_Q \mathbin{\&} M_L$
    \State $F_{\max} \gets F_{\min} \mathbin{|} \overline{M_L}$
    \State $i \gets \Call{LowerBound}{P,F_{\min}}$
    \State $j \gets \Call{UpperBound}{P,F_{\max}}$
    \State \Return $i<j$
\EndFunction
\end{algorithmic}
\end{algorithm}

\textbf{Estimating match potential.}
Given a query \(Q\) and a candidate segment \(c\) returned by the routing entry at \(p\), SARP computes the corresponding filter (i.e., region fingerprints) for \(Q\) and compares them with those of \(c\) in sequence order, starting with the first region, which starts at \(p\).
The comparison stops at the first mismatching field.
Because each fingerprint represents the prefix through the end of its region, a field mismatch proves that \(c\) and \(Q\) diverge before that region ends.
In contrast, a matching field does not guarantee that the prefixes match, because truncated fingerprints may collide.
Thus, if field \(j\) is the first mismatch, \(c\) can match at most $j\cdot\mathit{stride}-1$ blocks, counted from \(p\).
This value is the candidate's filter-derived match potential.

\textbf{Finding candidates with match potential at least \(L\).}
The filter comparison above gives a threshold test for selecting candidates
before segment retrieval, which avoids reading every candidate in a large
posting list. Let \(L\) denote a target match length, measured in blocks from
the routing point. For a candidate to have a filter-derived match potential of
at least \(L\), its filter needs to match the first
\(k=\left\lfloor L/\mathit{stride}\right\rfloor\) fields of the query filter.
Because these fields occupy the most significant bits, all filters sharing these \(k\) query fields fall within a contiguous interval in unsigned integer order. Since the posting list of each routing entry is \emph{sorted} by filter value, the corresponding candidates also form a contiguous range in the posting list.

SARP constructs the lower and upper bounds of this interval by fixing the first \(k\) fields to the query fields while setting all remaining bits to zero or one, respectively.
Specifically, let \(M_L\) retain the first \(k\) query fields and clear all remaining bits. We define

$$
F_{\min}(L)=F_Q\mathbin{\&}M_L,
\qquad
F_{\max}(L)=F_{\min}(L)\mathbin{|}\overline{M_L}.
$$

As \textsc{ExistsMatch} in \autoref{alg:find-max-potential} illustrates, every candidate whose filter-derived match potential is at least \(L\) has a filter value within
\([F_{\min}(L),F_{\max}(L)]\).
SARP uses binary search on the sorted posting list to find the first candidate whose filter is no smaller than \(F_{\min}(L)\). Candidates from this position may match \(L\) or more, until the filter value exceeds \(F_{\max}(L)\).
For example, \autoref{fig:filter-layout} uses the query fields
\(F_Q=12\mathbin{|}34\mathbin{|}56\mathbin{|}78\mathbin{|}\cdots\) with
\(\mathit{stride}=2\). For \(L=4\), \(k=2\), so \(F_{\min}(4)\) fixes the
first two fields to \(12\mathbin{|}34\), while \(F_{\max}(4)\) allows all
later fields to vary. The lower-bound and upper-bound probes therefore select
every filter with this prefix. SARP uses this range to restrict the
candidates considered for retrieval because exact verification requires loading
each selected segment. This avoids reading candidates whose filters already
show that they cannot reach the target match length.

\textbf{Finding the largest match potential.}
The \textsc{ExistsMatch} existence test is monotonic in \(L\): if no candidate has match potential of at least \(L\), no candidate can have match potential of at least any \(L'>L\), because a larger length requires matching a superset of the leading query fields.
\autoref{alg:find-max-potential} therefore binary-searches \([0,L_{\max}]\) for the largest \(L\) whose filter interval is nonempty.
We denote this maximum filter-derived match potential by \(L^*\).

If the posting list contains the three candidates shown in \autoref{fig:filter-layout},
their filter order is:
\[
\begin{array}{c|c|c}
\text{position} & \text{candidate} & \text{filter-field prefix} \\ \hline
0 & c_3 & 12\mathbin{|}34\mathbin{|}56\mathbin{|}9D\mathbin{|}\cdots \\
1 & c_1 & 12\mathbin{|}34\mathbin{|}A9\mathbin{|}\cdots \\
2 & c_2 & 12\mathbin{|}B7\mathbin{|}\cdots
\end{array}
\]
For \(L=4\), the two bound probes return \(i=0\) and \(j=2\), so
positions \([0,2)\) select \(c_3\) and \(c_1\) while excluding \(c_2\).
The first mismatch shown in \autoref{fig:filter-layout} gives these
candidates filter-derived match potentials of at most 7, 5, and 3 blocks,
respectively. At \(L=7\), only \(c_3\) matches the first three query fields,
so the interval is \([0,1)\). At \(L=8\), no filter matches the first four
fields, so the interval is empty. \autoref{alg:find-max-potential}
therefore returns \(L^*=7\) and considers \(c_3\) first for retrieval.
For a posting list containing \(n\) candidates, the binary search performs
\(O(\log L_{\max})\) pairs of bound probes, each taking \(O(\log n)\) time,
for total time \(O(\log L_{\max}\log n)\). In this example, it identifies the
relevant positions without scanning the full list.

\textbf{Filter tradeoff and accuracy.}
Under a fixed 64-bit budget, the number of fields \(N\) trades off the granularity of the match-potential bound against fingerprint width \(b=64/N\).
Increasing \(N\) creates more, smaller regions and therefore yields a finer-grained bound, but leaves fewer bits per fingerprint and increases the probability of collisions.
SARP uses \(N=8\) and \(b=8\) by default.

\subsection{Adaptive Routing-Point Pruning}
\label{sec:routing-point-pruning}

Adaptive routing-point refinement reduces large posting lists by adding routing points,
and candidate filtering limits segment retrievals by prioritizing candidates
with larger match potential. Neither mechanism reduces the routing state
introduced by refinement. Each added routing entry enlarges the in-memory
state that recovery restores. The remaining goal is therefore to remove
routing points that add little discrimination while keeping candidate
retrieval within the retrieval budget.

\begin{figure}[tbp]
    \centering
    \includegraphics[width=\columnwidth]{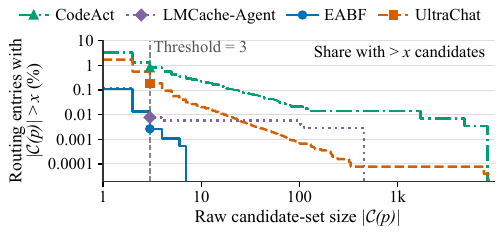}
    \caption{Distribution of raw candidate-set sizes
    \(|\mathcal{C}(p)|\) over routing entries under base placement,
    without refinement or pruning. Markers give the share of routing
    entries above \texttt{PruningThreshold} (3).}
    \label{fig:routing-candidate-set-size}
    \Description{A line chart with logarithmic axes shows, for each of
    CodeAct, LMCache-Agent, EABF, and UltraChat, the percentage of routing
    entries whose raw candidate-set size exceeds x. A dashed line marks the
    threshold of 3. Fewer than 1 percent of routing entries exceed the
    threshold on every workload. The CodeAct and UltraChat curves extend to
    thousands of candidates, whereas the EABF curve ends below 10.}
\end{figure}

\textbf{Observation and rationale.}
To determine when an additional routing point is useful, let
\(\mathcal{C}(p)\) denote the posting list returned by the routing entry
associated with routing point \(p\), before filtering, which forms the raw
candidate set.
\autoref{fig:routing-candidate-set-size} shows that
\(|\mathcal{C}(p)|\) is highly heterogeneous across routing points: some
posting lists contain many segment references, whereas many contain only one
or two. Across the four workloads, 99.2\% to 99.997\% of the routing entries
under base placement have posting lists no larger than the configured
\texttt{PruningThreshold} of 3. A posting list at or below this threshold is
small enough for the configured retrieval budget, so an additional routing
point would split an already manageable list while adding routing state. The
small fraction of lists above the threshold are the cases in which another
routing point can provide useful discrimination before candidate retrieval.
Thus, the posting-list size at the preceding retained routing point provides
a local signal: a small list favors pruning the next point, whereas a large
list justifies retaining it.

\textbf{Routing-point pruning.}
SARP applies pruning on top of the base placement policy rather
than replacing it. The first routing point of each segment is always
retained as its initial routing point. For every subsequent routing point
selected by the base placement policy, SARP examines the
posting-list size of the preceding retained routing point. If this list
contains no more than \texttt{PruningThreshold} candidates, SARP
prunes the new routing point and continues routing through the preceding
one. Otherwise, it retains the new routing point and materializes its
routing entry to provide additional discrimination.
\autoref{alg:routing-point-pruning} summarizes this decision.

Pruning does not extend any per-candidate filter's summarized range.
All candidates returned by the preceding routing entry can already be verified within
\texttt{MaxRetrieval}, so the retrieval budget and correctness still hold.

\begin{algorithm}[t]
\caption{Adaptive routing-point pruning.}
\label{alg:routing-point-pruning}
\begin{algorithmic}[1]
\State \textbf{state:} preceding retained routing point
       \(\mathit{previousPoint}\)
\State \textbf{parameter:} \(\texttt{PruningThreshold}\)

\Procedure{ShouldRetain}{$\mathit{ctx}, i$}
    \If{\(\neg\Call{BaseShouldPlace}{\mathit{ctx}, i}\)}
        \State \Return false
    \EndIf

    \If{\(\mathit{previousPoint}\) does not exist}
        \State \Return true
    \EndIf

    \If{\(|\mathcal{C}(\mathit{previousPoint})|
        \le \texttt{PruningThreshold}\)}
        \State \Return false
    \EndIf

    \State \Return true
\EndProcedure

\Procedure{OnRetain}{$p$}
    \State \(\mathit{previousPoint} \gets p\)
\EndProcedure
\end{algorithmic}
\end{algorithm}

\textbf{Routing-point rematerialization.}
A pruned routing entry may accumulate
additional candidates as new segments are inserted. Once its
posting-list size exceeds \texttt{PruningThreshold}, SARP
asynchronously materializes routing entries for the routing points that
were previously skipped. Constructing the per-candidate filters for these
entries requires the block IDs stored in the corresponding segments. A
background worker therefore retrieves the affected segments, re-applies
the base placement policy to recover their routing points, and constructs
the missing routing keys, routing entries, and per-candidate filters
outside the insertion critical path.

During rematerialization, the preceding routing entry remains authoritative,
and queries continue to use its existing filter-ranked selection subject
to \texttt{MaxRetrieval}. The rematerialized routing entries remain private
until all candidates in the preceding routing entry have been processed,
after which SARP publishes them atomically. Queries therefore
observe either the complete preceding routing entry or the complete
rematerialized routing entries, never a partially populated routing state.

\section{Implementation}
\label{sec:impl}

We implement SARP in C++20.
To separate index logic from storage-backend behavior, SARP accesses
segments through a common segment-store interface.
The same index therefore runs on RocksDB \cite{rocksdb},
WiredTiger \cite{mongodb2026wiredtiger}, and
BloomStore \cite{lu2012bloomstore}.

\textbf{Concurrent routing directory.}
The in-memory routing directory uses a oneTBB concurrent hash
map~\cite{onetbb-concurrent-hash-map} so that request threads can match
and insert prefixes concurrently.
The map's per-entry accessors let concurrent reads share a routing entry
while an update takes exclusive access.
Each read therefore observes either the old or the new complete posting list,
never a partially updated list.

\textbf{Block-ID verification.}
The segment store uses a columnar layout for post-retrieval verification.
It places all block IDs in one contiguous region and the corresponding block
metadata in a separate contiguous region.
After retrieving a segment, SARP scans the block IDs against the query
and batches these comparisons with AVX2~\cite{intel2012isa}.

\textbf{Routing-directory snapshots.}
Snapshots contain the routing points, posting lists, filters, and candidate
references needed to rebuild the in-memory routing directory.
The complete block IDs and block metadata remain in the durable segment store,
so they are not duplicated in the routing snapshot.
To publish a snapshot, SARP writes a temporary file, calls
\texttt{fsync} on that file, atomically renames it, and calls \texttt{fsync} on
the parent directory.
After a crash, this sequence exposes either the previous complete snapshot or
the newly published complete snapshot.

\section{Evaluation}
\label{sec:evaluation}

\begin{figure*}[t]
    \centering
    \includegraphics[width=\textwidth]{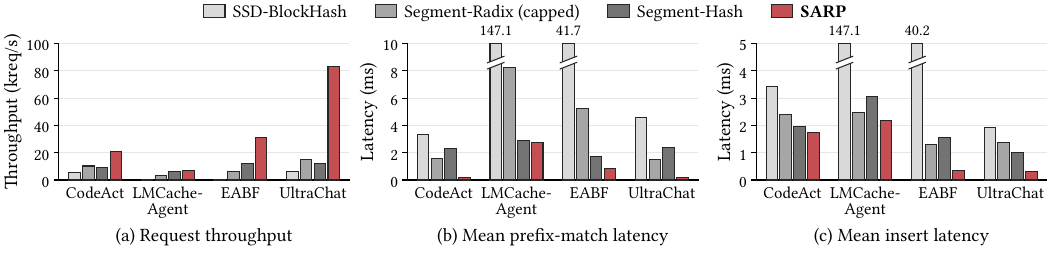}
    \caption{Index performance. (a) Request throughput. (b) Mean
    prefix-match latency. (c) Mean insert latency. Bars beyond the axis
    limit are cut by double slashes and labeled with their means.}
    \label{fig:eval-perf}
    \Description{Three grouped bar charts on linear axes compare
    SSD-BlockHash, Segment-Radix (capped), Segment-Hash, and SARP
    across CodeAct, LMCache-Agent, EABF, and UltraChat. SARP has the
    highest request throughput and lowest mean prefix-match latency on every
    dataset. Its mean insert latency is lower than that of Segment-Radix (capped)
    on every dataset.}
\end{figure*}

We conduct comprehensive evaluations to answer the following questions:

\begin{enumerate}[labelindent=0pt, leftmargin=*]
    \item \textbf{Index performance.} How does SARP compare with baselines under comparable token hit rate and recovery time?
    \item \textbf{Recovery efficiency.} How much does SARP reduce recovery time and the state rebuilt after a restart?
    \item \textbf{Prefix reuse.} How does the token hit rate of SARP's sparse routing compare with exact prefix matching?
    \item \textbf{Design effectiveness.} What are the effects of routing-point refinement, candidate filtering, and routing-point pruning on in-memory state and performance?
    \item \textbf{Space amplification.} What storage space amplification does segment packing introduce relative to the non-packing version (SSD-BlockHash)?
    \item \textbf{Multi-threading scalability.} How does SARP's performance scale as the number of request threads increases?
\end{enumerate}

\subsection{Experimental Setup}
\label{sec:eval-setup}

\textbf{Testbed.}
We use a server with two 28-core Intel Xeon Gold 6330 CPUs,
751\,GiB of DRAM, and a 2\,TB Samsung 980 PRO NVMe SSD with XFS.
The server runs Ubuntu 22.04 with Linux 6.8, and we compile all designs
with GCC 11.4.

\textbf{Baselines.}
On this testbed, we compare SARP with the following baselines:
\begin{itemize}[labelindent=0pt, leftmargin=*]
    \item \textbf{DRAM-BlockHash} stores one metadata entry per block
    in DRAM following vLLM's block-hash approach~\cite{kwon2023efficient,vllmPrefixCachingDocs}.

    \item \textbf{SSD-BlockHash} stores the same per-block entries in
    RocksDB.

    \item \textbf{Segment-Hash} packs metadata into segments and computes
    their keys directly from the query.
    On a miss, it retries at successively halved segment sizes down to
    one block to recover finer-grained matches.

    \item \textbf{Segment-Radix} stores metadata in segments and uses an
    in-memory radix tree for exact prefix matching.
    Its capped variant, \textbf{Segment-Radix (capped)}, limits in-memory
    routing state by spilling nodes to RocksDB.
\end{itemize}
DRAM-BlockHash serves as a reference for exact in-memory prefix matching and recovery,
while SSD-BlockHash exposes the storage access cost of a naive on-SSD index.
Segment-Hash and Segment-Radix share SARP's segment layout, allowing
us to evaluate sparse routing against the storage-access cost and memory cost of naive exact routing, respectively.

\textbf{Datasets.}
We use the four datasets listed in \autoref{tab:workloads}.
Among the agent datasets, CodeAct~\cite{wang2024codeact}
(\texttt{CodeActInstruct}) contains sessions in which agents write and run
code, while LMCache-Agent~\cite{lmcache2026agentic} contains longer agent
sessions.
Among the chat datasets, EABF~\cite{zhang2024extending} contains long
ShareGPT conversations, while UltraChat~\cite{ding2023ultrachat}
(\texttt{ultrachat}\allowbreak\texttt{\_200k}) contains short synthetic
chats.
We tokenize all four datasets with the Llama-2 tokenizer~\cite{touvron2023llama2}.

\textbf{Methodology.}
We simulate LLM requests from these datasets using only the index, without
LLM inference or KVCache data transfer.
For each turn, we first perform prefix matching on its context,
which includes the session's accumulated history.
We then append the turn's output to this context and insert
the resulting sequence into the index for reuse in the next turn.
We process sessions in round-robin order, completing the current turn across
sessions before advancing to the next turn.
To account for variation between runs, we repeat each experiment three times.

\textbf{Configuration.} For performance evaluation, we use 42 request threads and 14 threads for RocksDB itself. All SSD-backed designs share a per-dataset RocksDB memory budget equal to 10\% of SSD-BlockHash's uncompressed store size, split equally between the block cache and memtable. For CodeAct, LMCache-Agent, EABF, and UltraChat, these budgets are 5, 16, 29, and 362\,MiB, respectively. We use RocksDB 11.1.1~\cite{rocksdb}, with direct I/O and write-ahead log enabled. We use 16-token blocks and 64-block segments where applicable, with one SARP configuration across datasets: \(R=4\) base routing points per segment, 64-bit filter with 8-bit fields at a two-block stride, \(\tau=256\), \(\texttt{PruningThreshold}=3\), and \(\texttt{MaxRetrieval}=3\).

\subsection{Overall Performance}
\label{sec:eval-perf}

We compare SARP's performance with that of the baselines at comparable token hit rates and recovery times. Segment-Radix takes substantially longer to recover because it rebuilds its in-memory radix tree. We therefore limit its in-memory state by spilling radix-tree nodes to backing storage, reducing its recovery time to 0.96–1.96 times that of SARP.

\textbf{Overall throughput.}
Under this cap-selection procedure, request throughput measures completed
requests per second for requests that perform both prefix matching and
insertion. \autoref{fig:eval-perf}a shows that SARP achieves the
highest request throughput on all four datasets, outperforming SSD-BlockHash by
3.70 to 66.6\(\times\), Segment-Radix (capped) by 2.04 to
5.67\(\times\), and Segment-Hash by 1.15 to 6.78\(\times\).

\textbf{Prefix matching.}
To explain the aggregate throughput result, we first isolate prefix matching.
We report mean prefix-match latency for one complete prefix-matching invocation
across the four datasets. \autoref{fig:eval-perf}b shows that SARP
has the lowest mean prefix-match latency on every dataset, reducing latency by
16.6 to 53.9\(\times\) relative to SSD-BlockHash, by
3.0 to 7.7\(\times\) relative to Segment-Radix (capped), and by
1.05 to 12.1\(\times\) relative to Segment-Hash. LMCache-Agent
has the smallest reduction against both segment baselines, so the next
paragraph examines its access paths.

\begin{figure}[t]
    \centering
    \includegraphics[width=\columnwidth]{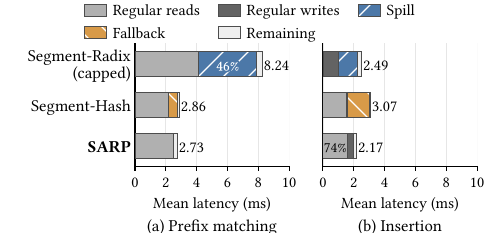}
    \caption{Latency breakdown. (a) Prefix matching. (b) Insertion. Stacks show reads, writes, spill I/O, fallback I/O, and Remaining.}
    \label{fig:eval-breakdown}
    \Description{Two horizontal stacked bar charts compare mean prefix-match
    latency and mean insert latency for Segment-Radix (capped), Segment-Hash,
    and SARP on LMCache-Agent. Spill and fallback components add to
    prefix-match latency for Segment-Radix (capped) and Segment-Hash, respectively.
    SARP has lower mean prefix-match and insert latency than both baselines.
    Regular reads dominate SARP's insert latency.}
\end{figure}

The breakdown measurements show that storage accesses account for 92 to
99\% of measured prefix-match latency on LMCache-Agent across the four
designs, making them the dominant cost. The stacked bars place routing,
hashing, decoding, and other unpartitioned time in \texttt{Remaining}.
SARP combines segment packing with sparse routing to reduce key-value
store reads, requiring 29.9 logical key-value store reads per prefix
match. These reads count requested key occurrences, including misses, batched
keys, and repeated requests, and do not equal physical SSD reads. SSD-BlockHash
keeps one on-SSD entry per block and therefore issues one logical key-value
store read for each requested key occurrence, reaching 1,840.2 logical key-value
store reads per prefix match and incurring higher prefix-match latency. Segment-Radix (capped)
spills radix-tree nodes to RocksDB. \autoref{fig:eval-breakdown}a
shows that spill reads and writes account for 46\% of its prefix-match
latency. Segment-Hash probes smaller segments when a larger segment is
insufficient. The same panel shows that fallback reads account for 22\%
of its prefix-match latency.

\textbf{Insert latency.}
To explain the other component of request throughput, we report mean insert
latency for adding the current context and turn output to the index for later
KVCache reuse. \autoref{fig:eval-perf}c shows that SARP has the
lowest mean insert latency on every dataset, reducing latency by 1.98 to
111\(\times\) relative to SSD-BlockHash, by 1.15 to
4.30\(\times\) relative to Segment-Radix (capped), and by 1.12 to
4.31\(\times\) relative to Segment-Hash.

On LMCache-Agent, SARP performs candidate selection and post-retrieval
verification during insertion to identify the reusable prefix before updating
the index. \autoref{fig:eval-breakdown}b shows that time spent on segment
retrievals accounts for 1.60\,ms of its 2.17\,ms mean insert
latency.
SSD-BlockHash reads and writes block-hash entries through the key-value
store, leaving its insert latency above SARP's. Segment-Radix (capped)
uses its exact radix tree to identify the stored prefix without repeated
segment retrievals. Segment-Hash incurs additional reads for post-retrieval
verification and fallback I/O during insertion, leaving its latency above
SARP's. These access paths explain the insert-latency ordering. SARP's margin
is smallest on CodeAct, 1.12\(\times\) over Segment-Hash. Because CodeAct is
the smallest dataset, its small RocksDB memtable budget causes write stalls
that delay insertion for both SARP and Segment-Radix
(\autoref{sec:eval-scalability-sensitivity}), so this narrow margin is not
specific to SARP.

\subsection{Recovery Time}

\label{sec:eval-recovery}
To compare recovery costs across index representations, we use one
recovery thread for SARP, DRAM-BlockHash, and uncapped Segment-Radix,
avoiding differences between parallel recovery implementations.
Given the limited dataset sizes, we scale each dataset to about one
billion unique index entries by replication,\footnote{Token IDs are offset
into disjoint ranges across dataset copies, so prefix sharing remains within
each copy. The resulting sharing structure is local to each copy and does
not scale with the number of copies.} counting one entry per
unique cached block.

\begin{figure}[tbp]
    \centering
    \includegraphics[width=\columnwidth]{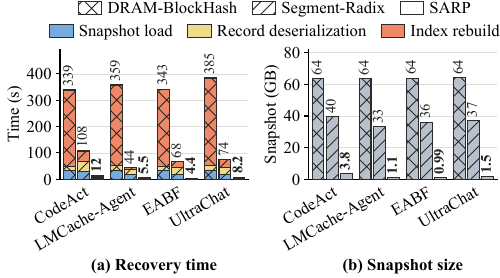}
    \caption{Recovery time and snapshot size (GB).}
    \label{fig:eval-recovery}
    \Description{Two side-by-side grouped bar charts compare
    DRAM-BlockHash, Segment-Radix, and SARP across four datasets.
    SARP has the smallest snapshot and shortest recovery time.
    Rebuilding dominates DRAM-BlockHash's recovery.
    Record deserialization and index rebuilding together take most of
    Segment-Radix's and SARP's recovery time.}
\end{figure}

\textbf{Recovery time.}
We report recovery time as the time from snapshot loading until the index is ready for serving requests. As \autoref{fig:eval-recovery}a shows, SARP's median recovery time is 4.42 to 12.4\,s across the four scaled datasets, a reduction by a factor of 8.0 to 15.3 relative to Segment-Radix's 43.6 to 108\,s. Relative to DRAM-BlockHash's 339 to 385\,s, SARP reduces recovery time by a factor of 27.3 to 77.5.

\textbf{Snapshot footprint.}
This recovery advantage reflects the smaller amount of index state that SARP restores. As \autoref{fig:eval-recovery}b shows, the snapshots restored by SARP occupy 0.99 to 3.78\,GB across the four datasets, only 2.7 to 9.5\% of Segment-Radix's snapshot size on the same dataset. Because block IDs and their metadata remain in the persistent segment store, SARP snapshots only its routing directory. Segment-Radix's snapshots, however, retain block-ID sequences for exact prefix matching even after unbranched paths are compressed into edges to reduce the node count. Among the datasets, CodeAct has SARP's largest snapshot. This is consistent with the ablation on the original datasets, where CodeAct's larger candidate sets are associated with a larger relative increase in routing state from refinement and a smaller reduction from pruning (\autoref{sec:design-breakdown}).

\textbf{Recovery time breakdown.}
To identify which stages account for SARP's shorter recovery time, we break recovery down into snapshot loading, record deserialization, and index rebuilding. As \autoref{fig:eval-recovery}a shows, SARP spends 1.93 to 7.82\,s on snapshot loading and record deserialization of its snapshots, compared with 38.0 to 67.9\,s for Segment-Radix across the four datasets. Index rebuilding then takes SARP 2.47 to 4.60\,s, compared with 5.68 to 39.8\,s for Segment-Radix. DRAM-BlockHash spends 289 to 332\,s rebuilding its table. The gap to SARP reflects the insertions needed to restore each index: SARP inserts 16.1 to 29.1 million routing entries, compared with about one billion block-hash entries for DRAM-BlockHash.

\subsection{Prefix Reuse}

We ask how bounded candidate retrieval affects KVCache reuse relative to exact
block-level matching and which retrieval budget provides a useful tradeoff.

\textbf{Hit rate.} Token hit rate is the fraction of arriving prompt tokens
that prefix matching reuses. \autoref{tab:eval-hit-loss} reports that
SARP loses 0.0030 to 0.029\% of the tokens matched by
exact matching across the four workloads, with the largest loss on CodeAct.

\begin{table}[tbp]
  \caption{Token hit rate and loss relative to exact matching.}
  \label{tab:eval-hit-loss}
  \small
  \begin{tabular}{lrrr}
    \toprule
    & \multicolumn{2}{c}{Token hit rate (\%)} & \\
    \cmidrule(lr){2-3}
    Workload & Exact & SARP & Loss (\%) \\
    \midrule
    CodeAct       & 87.02 & 87.00 & 0.029 \\
    LMCache-Agent & 96.69 & 96.68 & 0.0087 \\
    EABF          & 96.86 & 96.85 & 0.0030 \\
    UltraChat     & 73.82 & 73.82 & 0.0037 \\
    \bottomrule
  \end{tabular}
\end{table}

CodeAct has the largest hit-rate loss in \autoref{tab:eval-hit-loss}, which
points to candidate multiplicity as the source of the difference. CodeAct
combines a shared prompt with diverse alternatives, so one query can match
the shared prefix in many stored segments. \autoref{tab:eval-filter-funnel}
reports an average of 59.3 candidate segments per matching step for
CodeAct before candidate filtering. SARP then retrieves at most
\texttt{MaxRetrieval} candidates per step, 3 by default, for block-by-block
verification. When the segment with the longest match falls outside this
budget, verification ends with a shorter prefix.

\begin{figure}[tbp]
    \centering
    \includegraphics[width=\columnwidth]{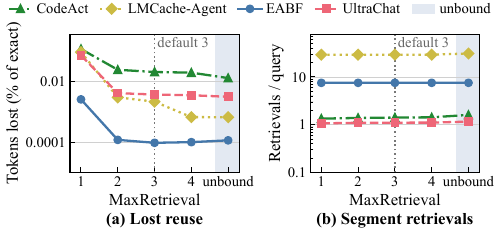}
    \caption{SARP sensitivity to \texttt{MaxRetrieval}. (a) Tokens
    lost relative to an exact oracle in the same run. (b) Segment retrievals
    per query at the key-value store interface, not physical SSD reads.
    ``unbound'' removes the cap.}
    \label{fig:eval-max-retrieval}
    \Description{Two line charts against MaxRetrieval from 1 to unbound,
    one line per workload. Tokens lost fall most sharply from 1 to 2,
    with much smaller changes at higher budgets. From 2 on, CodeAct loses
    the most. Segment retrievals per query remain nearly flat on
    LMCache-Agent, EABF, and UltraChat, and increase with the budget
    on CodeAct.}
\end{figure}

\textbf{Retrieval budget sensitivity.} \texttt{MaxRetrieval}=3 is a sweet spot
for the evaluated workloads: larger budgets add retrievals while recovering
little additional prefix reuse. \autoref{fig:eval-max-retrieval} shows this
tradeoff. Unlike \autoref{tab:eval-hit-loss}, which compares separate
concurrent runs, \autoref{fig:eval-max-retrieval} compares each query with
an exact oracle in the same run to isolate the effect of the budget. Token loss
ranges from 0.0025 to 0.114\% at a budget of 1 and from 0.00011 to
0.0127\% without a cap. Most loss reduction occurs from 1 to 2. On CodeAct, loss is 0.020\% at the
default budget of 3 with 1.41 segment retrievals per query. Increasing
the budget beyond 3 lowers loss to 0.0127\% but increases retrievals by
13.7\% to 1.60 per query. On the other workloads, the same
increase changes loss by at most 0.0015 percentage points.

\subsection{Design Breakdown}\label{sec:design-breakdown}

Adaptive routing-point refinement and filtering reduce candidates, while
adaptive routing-point pruning reduces routing state with a small increase in
segment retrievals. We evaluate these mechanisms in Design order.

\begin{figure}[tbp]
    \centering
    \includegraphics[width=\columnwidth]{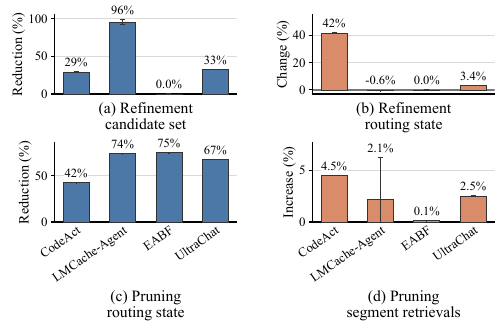}
    \caption{Effects of routing-point refinement and routing-point pruning. (a)
    Candidate-set reduction. (b) Refinement's routing-state change. (c)
    Pruning's routing-state reduction. (d) Pruning's segment-retrieval
    increase. Error bars show 95\% confidence intervals. The horizontal line
    marks no change.}
    \label{fig:eval-ablation}
    \Description{Four bar charts compare routing-point refinement and routing-point
    pruning across CodeAct, LMCache-Agent, EABF, and UltraChat. Refinement
    reduces candidate-set size and changes routing state. Pruning reduces
    routing state and increases segment retrievals.}
\end{figure}

\textbf{Adaptive routing-point refinement.}
Adaptive routing-point refinement reduces candidate-set size while increasing routing state.
\autoref{fig:eval-ablation}a shows that refinement reduces the candidate set
by 29\% on CodeAct, 96\% on LMCache-Agent, 0\% on EABF, and 33\% on UltraChat
relative to the corresponding configurations without refinement.
\autoref{fig:eval-ablation}b shows that refinement increases routing state by
42\% on CodeAct and 3\% on UltraChat, while leaving it essentially unchanged
on LMCache-Agent and EABF. CodeAct requests share an initial prompt but differ
later, so additional routing points are useful for reducing the resulting
candidate sets.

\textbf{Candidate filtering.}
Candidate filtering further narrows the remaining candidate set before exact
verification. \autoref{tab:eval-filter-funnel} shows that filtering reduces
the candidate count from 59.3 to 31.1 on CodeAct, from 1.78 to 1.66 on
LMCache-Agent, from 1.01 to 1.00 on EABF, and from 24.3 to 1.26 on UltraChat.
Its Actual Retrieved column gives the segment-retrieval amplification, the mean
number of segments retrieved at a routing entry, which is 1.05, 1.00, 1.00,
and 1.03, respectively. Retrieval stops once it
finds a candidate matching the filter-estimated length, so the number of
segment retrievals can be smaller than the number of candidates remaining
after filtering. This reduction is most visible when a routing entry starts
with many candidates, especially in CodeAct and UltraChat.

\begin{table}[t]
  \caption{Mean candidate counts before filtering, after filtering, and
  actually retrieved at a SARP routing entry.}
  \label{tab:eval-filter-funnel}
  \centering
  \small
  \begin{tabular}{lrrr}
    \toprule
    Workload & \shortstack{Before\\Filtering} &
    \shortstack{After\\Filtering} & \shortstack{Actual\\Retrieved} \\
    \midrule
    CodeAct       & 59.3 & 31.1 & 1.05 \\
    LMCache-Agent & 1.78 & 1.66 & 1.00 \\
    EABF          & 1.01 & 1.00 & 1.00 \\
    UltraChat     & 24.3 & 1.26 & 1.03 \\
    \bottomrule
  \end{tabular}
\end{table}

\textbf{Adaptive routing-point pruning.}
Under the retrieval budget, adaptive routing-point pruning trades lower routing
state for a small increase in segment retrievals.
\autoref{fig:eval-ablation}c shows that pruning reduces routing state by
42\% on CodeAct, 74\% on LMCache-Agent, 75\% on EABF, and 67\% on UltraChat.
\autoref{fig:eval-ablation}d shows that segment retrievals increase by
4.5\% on CodeAct, 2.1\% on LMCache-Agent, 0.1\% on EABF, and 2.5\% on
UltraChat.

\subsection{Space Amplification}
\label{sec:eval-space}

\begin{table}[b]
  \caption{Storage space and logical block-metadata entries for the packed and non-packing versions.}
  \label{tab:packing-amplification}
  \centering
  \scriptsize
  \setlength{\tabcolsep}{3pt}
  \begin{tabular}{@{}lrrr@{\hspace{6pt}}rrr@{}}
    \toprule
    & \multicolumn{3}{c}{Storage (MiB)}
    & \multicolumn{3}{c}{Logical entries} \\
    \cmidrule(lr){2-4}\cmidrule(lr){5-7}
    Workload & Non-packing & SARP & Ratio & Non-packing & SARP & Ratio \\
    \midrule
    CodeAct       & 17.5   & 14.7   & 0.844 & 440,235    & 622,782    & 1.415 \\
    LMCache-Agent & 59.7   & 35.6   & 0.597 & 1,552,387  & 1,584,809  & 1.021 \\
    EABF          & 108.3  & 62.6   & 0.578 & 2,800,814  & 2,802,755  & 1.001 \\
    UltraChat     & 1,387.2& 930.4  & 0.671 & 17,843,596 & 17,956,521 & 1.006 \\
    \bottomrule
  \end{tabular}
\end{table}

To quantify the storage cost of segment packing, we compare SARP with the
non-packing version, SSD-BlockHash, at the storage-space and logical
block-metadata levels. \autoref{tab:packing-amplification} shows that SARP
uses 0.58 to 0.84 times the non-packing version's storage space across the
four workloads, which corresponds to a 15.6\% to 42.2\% reduction. The
1024-token segment
layout still carries 1.001 to 1.415 times as many logical block-metadata
entries, or 0.069\% to 41.5\% more entries, because shared-prefix metadata
can be repeated across alternatives. The storage space is smaller
because each segment record amortizes the key, header, and storage-engine
metadata that the key-value store keeps per record across many block-metadata
entries. The logical ratio therefore captures duplicated block metadata,
whereas the storage-space ratio also captures the reduced per-record storage
overhead.

\subsection{Scalability}
\label{sec:eval-scalability-sensitivity}

\begin{figure}[t]
  \centering
  \includegraphics[width=\columnwidth]{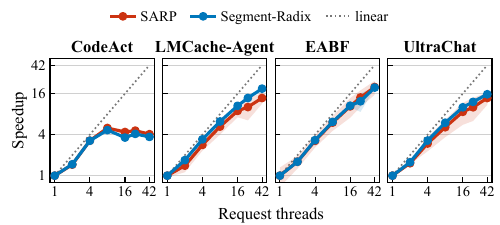}
  \caption{Request throughput versus request threads, normalized to one request
  thread.}
  \label{fig:eval-sens-threads}
  \Description{Four panels in one row plot normalized request throughput against
  request threads for CodeAct, LMCache-Agent, EABF, and UltraChat. SARP and
  Segment-Radix increase almost linearly through 42 request threads on the
  latter three workloads. Both methods peak near eight request threads on
  CodeAct and decline at 42 threads.}
\end{figure}

\autoref{fig:eval-sens-threads} plots request
throughput normalized to one request thread. SARP's request throughput
increases almost linearly through 42 request threads on LMCache-Agent, EABF,
and UltraChat. It reaches 13.8$\times$, 20.0$\times$, and 13.8$\times$,
respectively, on those workloads. CodeAct is the exception: SARP and
Segment-Radix both peak at eight request threads and decline at 42 threads.
On CodeAct, the small RocksDB memtable budget causes write stalls that grow
with the number of request threads and limit both designs, which explains the
CodeAct plateau.

\section{Related Work}
\label{sec:related-work}

\textbf{KVCache indexes.}
vLLM, SGLang, Mooncake, NVIDIA Dynamo, llm-d, and MemServe use in-memory
block-hash or radix-tree indexes for prefix matching
~\cite{kwon2023efficient,zheng2024sglang,mooncakeStoreDocs,dynamoRouterDocs,
llmdKVIndexerDocs,hu2024memserve}. They keep in-memory state for every cached
block, which supports exact longest-prefix matching without storage
retrievals, but their restart recovery grows with the cache. Dynamo's Flash
Indexer probes selected positions and scans skipped positions after a
mismatch~\cite{dynamo-flashindexer}. This sparse probing reduces matching
work, but its in-memory index still holds an entry for every cached block.
SARP instead makes the in-memory state itself sparse to shorten recovery. It
keeps routing entries only at selected routing points, leaves block IDs and
block metadata in on-SSD segments, and restores only this routing directory
after a restart, at the cost of segment retrievals during prefix matching.

\textbf{Fast recovery of in-memory state.}
In-memory storage systems shorten recovery by restoring their state faster.
RAMCloud scatters backup data across many disks so that many servers can
recover a crashed server's DRAM contents in
parallel~\cite{ongaro2011ramcloud}. Zheng et al.\ parallelize logging,
checkpointing, and recovery across cores to restore an in-memory
database~\cite{zheng2014silor}. Redis and Mooncake reload a snapshot of the
whole index~\cite{redisPersistenceDocs,mooncakeStoreDocs}. These systems speed
up restoring the same state. SARP instead reduces the state that recovery
restores, keeping only the sparse routing directory in memory while block IDs
and block metadata remain in durable segments.

\textbf{In-memory prefix indexes.}
General prefix indexes such as PATRICIA, ART, HOT, and zip-tries compress
paths, nodes, or discriminating bits while retaining the distinctions needed
to choose a branch~\cite{morrison1968patricia,leis2013art,binna2018hot,ziptries}.
They therefore support exact longest-prefix navigation in DRAM. SARP
keeps only routing points in DRAM. One routing entry can name several
candidate segments, so prefix matching resolves omitted distinctions with
segment retrievals and block-ID comparison. A retrieval cap can
leave the longest candidate unexamined, but SARP reuses no KVCache
block until post-retrieval verification confirms its block-ID match.

\textbf{On-storage prefix indexes.}
Prefix B-trees, String B-trees, and their successors navigate ordered strings
through pages on storage~\cite{bayer1977prefix,ferragina1999stringbtree,
bender2006cacheoblivious,luccio2007stringaccess,clark1996secondarypat}.
Their ordered keys let a prefix match narrow the search through an ordered
range, while the search traverses index pages before reaching the associated
records. SGLang-LSM applies this key-ordering approach to KVCache with
token-sequence keys in an on-storage LSM tree~\cite{sglang-lsm}. SARP
uses fixed-size chained block IDs instead. These IDs preserve prefix identity
but not lexicographic order, so one routing key can name several candidate
segments rather than one ordered range. SARP selects those candidates
in its DRAM routing directory, retrieves and verifies the segments, and
restores the directory after a restart.

\textbf{Sparse indexes for deduplication.}
DDFS groups adjacent backup-segment descriptors so one metadata read loads
neighbors~\cite{zhu2008ddfs}. Sparse Indexing samples chunk hashes to select a
bounded set of manifests, then verifies exact matches after loading
~\cite{lillibridge2009sparse}. SARP applies this sampled-routing idea
to a stricter contract: KVCache reuse requires every earlier block to match,
so it ranks candidates by possible match length and retains routing points at
segment starts. Refinement and pruning control posting lists, while filters
rank segments before retrievals.

\section{Conclusion}

Persistent KVCache indexes trade recovery state against prefix-matching cost.
SARP (Sparse Adaptive Routing and Packing) addresses this tradeoff by keeping block IDs and their metadata in
on-SSD segments while retaining only selected routing points and candidate
references in memory. Segment retrievals verify matches exactly, while
adaptive routing-point refinement, filtering, and pruning balance candidate discrimination,
retrieval cost, and recovery state. Across four multi-turn chat and agentic
workloads, SARP recovers 8.0$\times$ to 77.5$\times$ faster than the in-memory
index baselines and delivers 1.15$\times$ to 66.6$\times$ the request
throughput of the on-SSD index baselines, while reusing only 0.0030\% to
0.029\% fewer tokens than exact prefix matching. These results show that a sparse persistent index can keep recovery fast while preserving high prefix-matching performance.

\bibliographystyle{ACM-Reference-Format}
\bibliography{references}

\end{document}